\documentclass[aps,prd,amsmath,longbibliography,twocolumn]{revtex4-1}

\usepackage[dvips]{graphicx}
\usepackage{latexsym,epsfig,bm,psfrag,subfigure}
\usepackage{color}         
\usepackage[bookmarksnumbered,bookmarksopen,colorlinks,citecolor=blue,linkcolor=blue]{hyperref} %

\newcommand{\eq}[1]{Eq.~(\ref{#1})}

\newcommand{\bfnabla}{\mbox{\boldmath$\nabla$}}
\def\bea{\begin{eqnarray}}
\def\eea{\end{eqnarray}}
\def\g{\gamma}\def\t{\tau}\def\l{\lambda}\def\s{\sigma}\def\o{\omega}
\def\bfp{{\bf  p}}
\def\la{\langle}\def\ra{\rangle}\def\d{\delta}\def\k{\kappa}\def\G{\Gamma}\def\z{\zeta}
\def\o{\omega}\def\bfb{{\bf b}}
\def\bfv{{\bf v}}\def\r{\rho}
\def\bfx{{\bf x}}
\def\bfP{{\bf P}}\def\wt{{2\k^2\over P^+}t}\def\bfQ{{\bf Q}}\def\bfb{{\bf b}}
\newcommand{\half}{{\frac{1}{2}}}\newcommand{\mbf}[1]{\mathbf{#1}}
\begin{document}



\title{NT@UW-22-03\\
Color Transparency and Light Front Holographic QCD}



\author{Gerald A. Miller}
\email{miller@uw.edu}
\affiliation{Department of Physics, University of Washington, Seattle, WA \ \ 98195, USA}

\date{\today}

\begin{abstract}
Color transparency, the reduction of initial- or final-state interactions in coherent nuclear processes, is a natural prediction of QCD provided that small-sized or point-like configurations (PLCs) are responsible for  high-momentum transfer, high-energy,  semi-exclusive processes. I use the FMS criteria for the existence of PLCs to show that the wave functions of light front holographic QCD, as currently formulated, do not contain a PLC.
  \end{abstract}




\maketitle

Color transparency is the amazing prediction of QCD that initial- and final-state interactions are reduced in high-momentum transfer, high-energy coherent reactions~\cite{Frankfurt:1994hf,Jain:1995dd,Ashery:2006zw,Dutta:2012ii}. 
Strong interactions are strong: when hadrons hit nuclei they generally break up the nucleus or themselves. Indeed, a well-known semi-classical formula states that the intensity of a beam of hadrons falls exponentially with the penetration distance through nuclei.  This effect is known as absorption.
It is remarkable that  QCD indicates 
 that, under certain specific conditions, the strong interactions can effectively be turned off and hadronic systems can move freely through a nuclear medium.\\

This phenomenon is based on three requirements:
\begin{itemize}
\item High momentum transfer coherent reactions are dominated by  point-like color-singlet components of the struck hadron wave function, denoted as PLCs. This statement was a prediction intialy based on perturbative QCD (pQCD). For example, early pQCD calculations~\cite{Farrar:1979aw,Efremov:1979qk,Lepage:1979za,Lepage:1979zb,PhysRevD.21.1636,DUNCAN1980159,PhysRevD.21.1636},
 of the pion elastic electromagnetic form factor (see Fig.~1a) were interpreted \cite{Mueller:1982bq} in the following manner:
A high-momentum, $Q$, photon hits one of the partons that  greatly increases the relative momentum to $Q$. The system can only stay together only by exchanging a gluon carrying that momentum. That gluon has a range of only  $1/Q$ so that the partons must be close together, making a small-sized system or point-like configuration (PLC). 
 The validity of  pQCD for computation of electromagnetic  form factors is a sufficient but not necessary condition for involvement of s PLC
 \cite{Frankfurt:1993es} .\\

\item Small objects have small cross sections. It has been widely reported that the imaginary part of the forward scattering amplitude, $f$ of 
a rapidly moving color singlet  object is proportional to the square of the transverse separation,  $b$, between  positive and. negative color charges.
The  dominant  contribution to $f$, at high energies  is given by  two-gluon exchange \cite{Low:1975sv,Nussinov:1975mw,Gunion:1976iy}, and the  remarkable feature is that, in the limit that $b$ approaches 0,  $f$ vanishes because color-singlet point particles do not exchange colored gluons. This feature is expressed concisely as $\lim_{b\to0}\s(b^2)\propto b^2.$  This reduced interaction, caused by interference between emission by quarks of different colors in coherent processes, is the basic ingredient behind QCD factorization proofs, is widely used \cite{Donnachie:2002en} and not questioned.\\

\item  A PLC, once created, will expand as it moves. This is because a PLC is not an eigenstate of the Hamiltonian. The expansion effect is diminished if the PLC moves with sufficiently high momentum.\\

\end{itemize}
If all three requirements are satisfied for a given coherent process the effects of color transparency will be evident.\\

The second and third items are  based on many calculations, many experiments and basic principles of quantum mechanics. The interesting dynamical question is the validity or lack thereof of the idea the PLCs are  involved. This question is intimately connected with the origin of hadronic electromagnetic form mechanism at high momentum transfer. In the perturbative QCD mechanism, the large  momentum transfer is taken up by the exchange of high-momentum gluons. For this to occur, all of the partons must be at the same transverse spatial location. 
Although it is natural to suppose that PLCs dominate  coherent high-momentum transfer
processes, it is far from obvious that this is the case~\cite{PhysRevD.21.1636,DUNCAN1980159,PhysRevD.21.1636,Feynman:1973xc}.  For example, 
 the momentum transfer can involve a single quark of high momentum. This is the Feynman mechanism, see Fig.~1, (first stated by Drell \& Yan \cite{Drell:1969km}). The spectator system is not required to shrink to a small size and color transparency effects involving protons would not be expected to occur. In this picture, a reasonably valid  relation between elastic and deep inelastic scattering is obtained. See also~\cite{West:1970av}. A more recent example  of a process that favors the Feynman mechanism appears in~\cite{PhysRevD.58.114008}. See also~\cite{ Burkardt:2003mb}. Note that
Feynman remarked, ``if a system is made of 3 particles, the large $Q^2$  behavior depends not on the singularity when just two come together, but rather when all three are on top of one another". Furthermore, ``such pictures are too simple and inadequate".\\

 \begin{figure}[h] \label{Diags}

		\includegraphics[width=.5\textwidth]{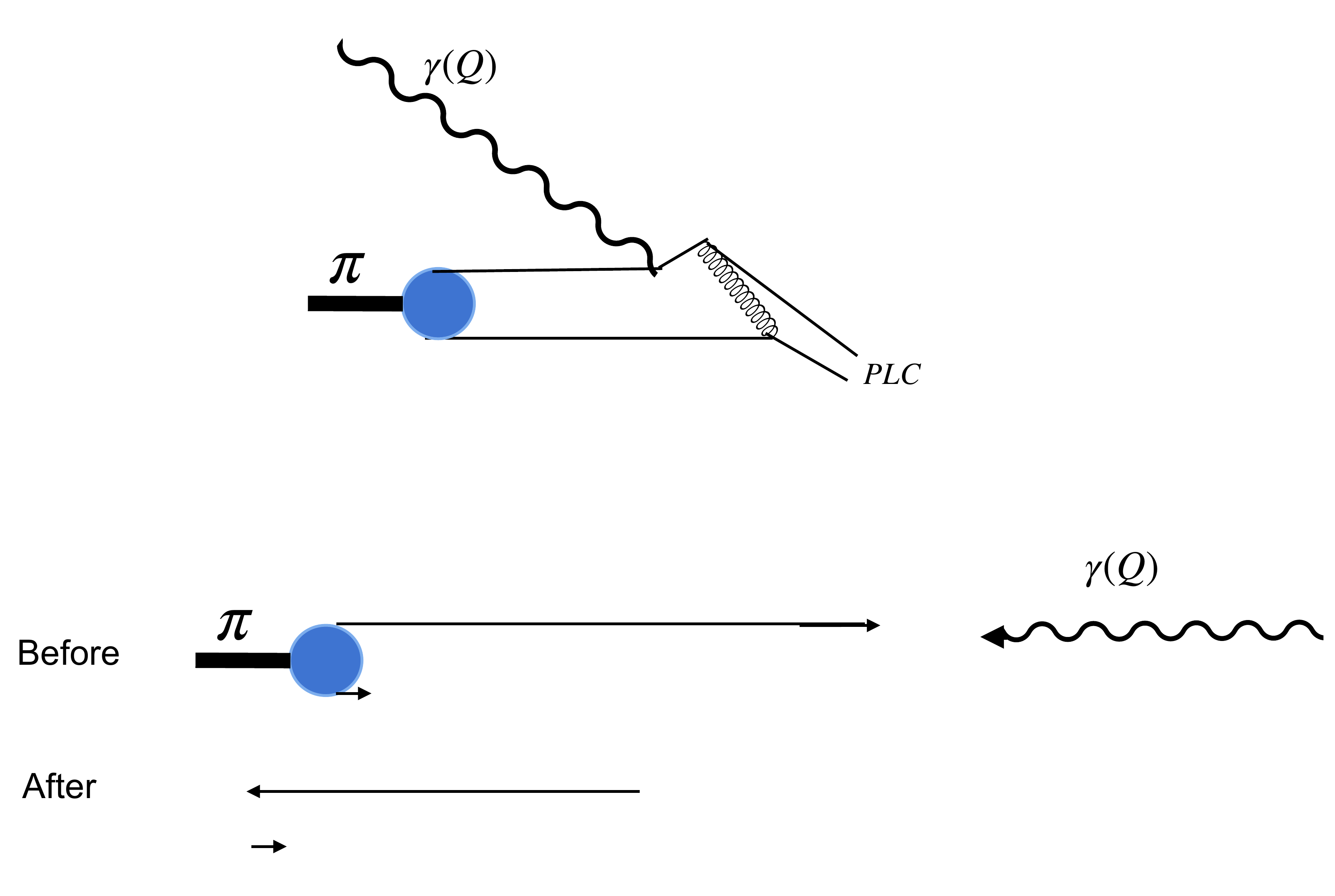} 
	      \caption{ High momentum transfer reaction mechanisms. Top picture: a pQCD mechanism. Other diagrams of the same order are not shown. Middle picture: Initial state in the Feynman mechanism. Bottom picture: final state in the Feynman mechanism. The final state has a good overlap with the turned around version of the initial state.}
      \end{figure}

 Models  of Generalized Parton Distributions (GPDS)~\cite{ Ji:2004gf} access both the longitudinal and transverse structure of nucleons so that measurements thereof can distinguish the different mechanisms. It has been said~\cite{Diehl:2003ny} that GPDs parameterize soft dynamics akin to the Feynman mechanism.   Specific models of GPDs, for example~\cite{deTeramond:2018ecg} also favor the Feynman mechanism.  More generally, a review of the  history  teaches us ~\cite{Belitsky:2005qn} that there are only two proposals for the mechanism responsible for  high momentum elastic reactions. \\
 
Frankfurt, Miller \& Strikman (FMS) introduced~\cite{Frankfurt:1993es} a criteria to determine whether or not a given model of a hadronic wave function admits the existence of a PLC. They found that a PLC could arise from non-perturbative effects as well as from perturbative  QCD. The aim of this paper is to use the FMS criteria to see if  the relativistic light-front wave functions obtained from  light front holographic QCD (see the review \cite{Brodsky:2014yha}) admit the existence of a PLC. \\

The first step is to discuss the FMS criterion. 
 The idea is that a PLC is originated via a  hard probing  interaction  $T_H$ involving nucleons initially bound in a nucleus.
The  soft interactions between the PLC and the surrounding medium  are  proportional to the square of the transverse separation distance \cite{Low:1975sv,Nussinov:1975mw,Gunion:1976iy} between constituents, $b^2=\sum_{i< j}(b_i-b_j)^2,$ where the constituents are labelled $i,j\cdots$ in first-quantized notation.  Consider a high-momentum transfer process on a nucleon. Denote  the initial nucleon  state as $|\psi(p)\rangle$ and the final state as $|\psi(p+q)\rangle$. Then represent the high momentum transfer operator as $T_H(q)|\psi\rangle$.  With this notation the form factor $F(Q^2)$ is given by the matrix element
\bea F(Q^2)=\la\psi(p+q)|T_H(q)|\psi(p)\rangle.\eea
The key question is whether the state $T_H(q)|\psi(p)\ra$ is a PLC that does not interact with the medium. The interaction with the surrounding medium is proportional to $b^2$, the square of the transverse separation between a struck parton and the remainder of the system. The first-order term in the interaction is proportional to  the matrix element $ \la\psi(p+q)|b^2 T_H(q)|\psi(p)\rangle$. This term is small if the operator $T_H$ produces a PLC. The relevant comparison is with the form factor $F(Q^2) $ that is the process amplitude in the absence of final state interactions. Thus FMS defined the quantity $b^2(Q^2)$ as
\bea b^2(Q^2)={\la\psi(p+q)|b^2T_H(q)|\psi(p)\rangle\over \la\psi(p+q)|T_H(q)|\psi(p)\rangle}\equiv{F_{b^2}(Q^2)\over F(Q^2)}\eea
If $b^2(Q^2)=b^2(0)$   final state interactions of normal magnitudes occur. If $b^2(Q^2)$ drops with increasing values of $Q^2$, then the model wave function is said to admit the existence of a PLC.\\

Now turn to evaluating $b^2(Q^2) $
for wave functions obtained from  holographic  techniques used  represent relativistic light front wave functionst, as discussed in the review~\cite{Brodsky:2014yha}.
I briefly discuss  that approach. Light-front quantization is a relativistic, frame-independent approach   to describing  the constituent structure of hadrons. The assumed simple structure of the light-front (LF) vacuum allows a  definition of the partonic content of a hadron in QCD and of hadronic light-front  theory~\cite{Brodsky:1997de}.  The spectrum and light-front wave functions of relativistic bound states are obtained in principle from the eigenvalue equation  $H_{LF} \vert  \psi \rangle  = M^2 \vert  \psi \rangle$ that becomes an infinite set of coupled integral equations for the LF components.  
 This provides a quantum-mechanical probabilistic interpretation of the structure of hadronic states in terms of their constituents at the same light-front time  $x^+ = x^0 + x^3$, the time marked by the front of a light wave~\cite{Dirac:1949cp}.      The matrix diagonalization~\cite{Brodsky:1997de} of the frame-independent  LF Hamiltonian eigenvalue equation in four-dimensional space-time has not  been achieved, so    other methods and approximations ~\cite{Brodsky:2014yha}  are needed to understand the nature of relativistic bound states in the strong-coupling regime of QCD.\\

To a first semiclassical approximation, where quantum loops and quark masses are not included, the relativistic bound-state equation for  light hadrons can be reduced to an effective LF Schr\"odinger equation. The technique is to  identify the invariant mass of the constituents  as a key dynamical variable, 
 $\zeta$, which measures the separation of the partons within the hadron at equal light-front  time~\cite{deTeramond:2008ht}.  Thus,  the multi-parton problem  in QCD is reduced, in   a first semi-classical approximation, to an effective one dimensional quantum field theory by properly identifying the key dynamical variable.  In this approach the complexities of the strong interaction dynamics are hidden in an effective potential. \\

It is remarkable that in the semiclassical approximation described above,  the light-front Hamiltonian  has a structure which matches exactly the eigenvalue equations in AdS space~\cite{Brodsky:2014yha}.  This offers the  possibility to explicitly connect  the AdS wave function $\Phi(z)$ to the internal constituent structure of hadrons. In fact, one can obtain the AdS wave equations by starting from the semiclassical approximation to light-front QCD in physical space-time.   This connection yields a  relation between the coordinate  $z$ of AdS space with the impact LF variable $\zeta$~\cite{deTeramond:2008ht},  thus giving  the holographic variable $z$ a precise definition and intuitive meaning in light-front QCD.\\

Light-front  holographic methods  were originally introduced~\cite{Brodsky:2006uqa,Brodsky:2007hb} by matching the  electromagnetic current matrix elements in AdS space~\cite{Polchinski:2002jw} with the corresponding expression  derived from light-front  quantization in physical space-time~\cite{Drell:1969km,West:1970av}. It was also shown that one obtains  identical holographic mapping using the matrix elements of the energy-momentum tensor~\cite{Brodsky:2008pf} by perturbing the AdS metric  around its static solution~\cite{Abidin:2008ku}, thus establishing a precise relation between wave functions in AdS space and the light-front wave functions describing the internal structure of hadrons.\\

The light front wave functions that arise out of this light front holographic approach provide a new way to study old problems that require the  use of relativistic-confining quark models. The study of  the existence of a PLC by evaluating $b^2(Q^2)$  is an excellent example of such a problem. \\

I evaluate two examples. The first~\cite{Brodsky:2007hb} is an early representation of the pion wave function as a $q\bar q$ system.
\bea \psi(x,b)={\k\over \sqrt{\pi}}\sqrt{x(1-x)}e^{-b^2\k^2x(1-x)/2}.\label{wf}\eea
I use the normalization $1=\int dxd^2b|\psi(x,b)|^2$ throughout this paper. The wave function of \eq{wf} appears in many models.
The form factor is given by
\bea F(Q^2)=\int dxd^2b e^{i \bfQ  \cdot\bfb(1-x)}|\psi(x,b)|^2,\eea and evaluation yields
\bea F(Q^2)=1-e^{Q^2/4\k^2}\G(0,Q^2/4\k^2),\eea
with $\G$ being the incomplete Gamma function. This form factor falls asymptotically as $\sim 1/Q^2$, providing a nice example of how a non-perturbative wave function can yield a power-law falloff. The limit for $Q^2\to0 $ is also interesting:
\bea 
\lim_{Q^2\to0}F(Q^2)=1 +(\g_E +\log(Q^2/4\k^2))Q^2/4\k^2+\cdots\eea
because of the appearance of the $\log$.\\

The quantity $F_{b^2}(Q^2)$ is obtained by inserting a factor $b^2$ into the integrand:
\bea& F_{b^2}(Q^2)=\int dxd^2b \,b^2\, e^{i \bfQ  \cdot\bfb(1-x)}|\psi(x,b)|^2\nonumber\\
&=-\nabla_Q^2\int {dx\over (1-x)^2}d^2b  e^{i \bfQ  \cdot\bfb(1-x)}|\psi(x,b)|^2\eea
Observe that $F_{b^2}$ is not simply obtained by differentiating the form factor with respect to $Q^2$. This is because of the factor $1-x$ that appears in the exponential function.
Evaluation of the integral over the transverse coordinates yields
\bea 
F_{b^2}(Q^2)={1\over \k^2}\int {dx \,e^{-{Q^2\over 4\k^2}}\over x^2(1-x)}(x+{Q^2\over 4\k^2}(1-x)).
\eea  
The value of $F_{b^2}(Q^2)$ is infinite for all values of $Q^2$ because of the divergence (related to the $\log Q^2$ term in the form factor) as $x$ approaches unity. This shows that the simple wave function of~\cite{Brodsky:2007hb} is not suitable  for use in evaluating high energy forward cross sections for pion-nucleus interactions.

 The model given in \eq{wf} is very simple, so  the next step is to use the universal light front wave functions of Ref.~\cite{deTeramond:2018ecg}.
   This work presented a universal description of generalized parton distributions obtained from Light-Front Holographic QCD, and I use their models for  light-front wave functions, presented  as  functions  of the number  $\t$ of constituents of a   Fock space component.   Regge behavior at small $x$ and inclusive counting rules as $x\rightarrow1$ are incorporated. Nucleon and pion   valence quark distribution functions  are obtained in precise agreement with global fits.  The model is described by the  
quark distribution  $q_\tau(x)$ and the profile function $f(x)$ with
\bea \label{qx} &
q_\tau(x) = \frac{1}{N_\tau} \big(1- w(x)\big)^{\tau-2}\, w(x)^{- \half}\, w'(x), \\&
 \label{fax}
f(x) =   \frac{1}{4 \lambda}\left[  (1-x) \log\left(\frac{1}{x}\right) + a (1 - x)^2 \right],
 \eea
and $w(x) = x^{1-x} e^{-a (1-x)^2}.$

The value of the universal scale $\lambda$ is fixed from the $\rho$ mass: $\sqrt{\lambda} = \kappa = m_\rho/ \sqrt{2} = 0.548$ GeV~\cite{Brodsky:2014yha,Brodsky:2016yod}. The   flavor-independent parameter  $a = 0.531 \pm 0.037$.   The $u$ and $d$ quark distributions of the proton  are given by  a linear superposition of $q_3$ and $q_4$ while those of the pion are obtained from $q_2$ and $q_4$.
Ref.~\cite{deTeramond:2018ecg} also presents
the universal light front wave function: 
\bea \label{LFWFb}
\psi_{\rm eff}^{(\tau)}(x, \mbf{b} ) = \frac{1}{2 \sqrt{\pi}} \sqrt{\frac{q_\tau(x)}{f(x)}} 
 (1-x) e^{ - \frac{(1-x)^2 }{8 f(x) } \, \mbf{b}^2},
\eea
in the transverse impact space representation. The transverse coordinate $b$ again represents the relative distance between a struck parton and the spectator system.\\

The form factor for a given value of $\t$ is given by
\bea F^{(\t)}(Q^2)=\int dx q_\t(x)e^{-Q^2 f(x)}, \eea
and $F^{(\t)}_{b^2}{Q^2}$, obtained by inserting a factor $b^2$ into the above integrand, is given by:
\bea &F^{(\t)}_{b^2}{Q^2}=\int dx {q_\t(x)\over 4f(x)}e^{-Q^2 f(x)}(1-{Q^2\over 16f(x)}(1-x)^2).\nonumber\\&\eea
 
 Consider first the case of $\t=2$.
 The use of \eq{qx} and \eq{fax} shows that
 \bea
 \lim_{x\to1}{q_2(x)\over f(x)}={4\l\over (1-x)}+\cdots.\eea
 Thus the same divergence that haunted the wave function of \eq{wf} reappears for the more sophisticated $\t=2$ wave function of
 \cite{deTeramond:2018ecg}
 The functions $b^2_\t(Q^2)= F^{(\t)}_{b^2}{Q^2}/F^{(\t)}{Q^2}$ for $\t=3,4$ are shown in Fig.~2.
  \begin{figure}[h]
\includegraphics[width=5.398cm,height=4.6cm]{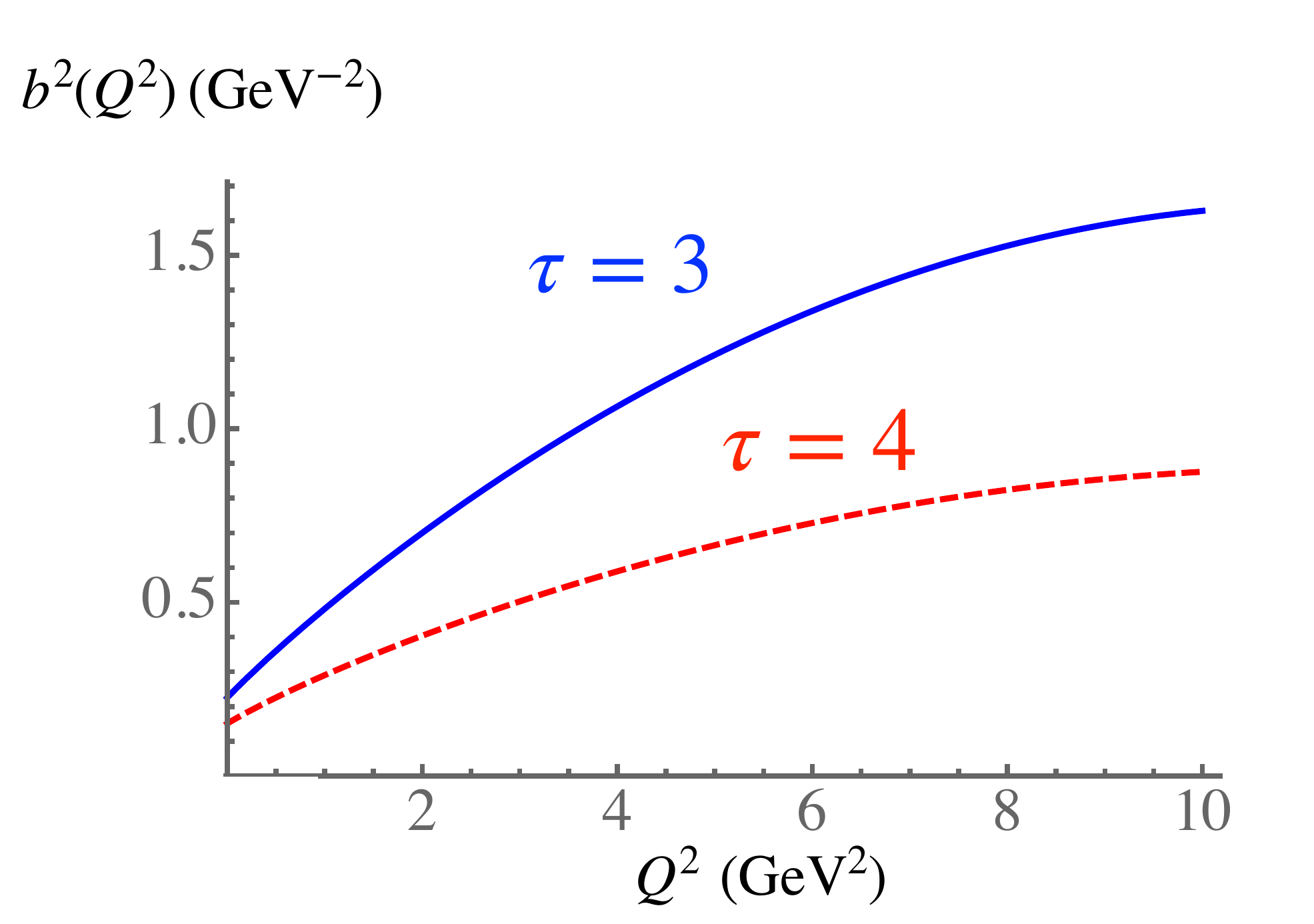}
 \caption{$b^2_\t(Q^2)$ in units of GeV$^{-2}$. The numbers refer to the value of $\t$,  the number of constituents in the Fock state.}\label{bsqz}\end{figure}\\
Observe that $b^2_\t(Q^2)$ rises with increasing values of $Q^2$, so that these wave functions do not admit the existence of PLCs. Furthermore, observe the surprising effect that constituents with larger number of partons have smaller values of $b^2(Q^2)$ and so  interact less strongly with a surrounding medium.\\

The summary of this work is that light front holographic wave functions do not contain a PLC, so  they do not predict the appearance of color transparency no matter how large the value  of $Q^2$.  Adding a perturbative QCD tail to the momentum-space wave function
could change this result. The wave functions of light front holographic QCD are suitable for describing the soft dynamics involved the time evolution of a wave packet and were used~\cite{Caplow-Munro:2021xwi} to interpret the  recent striking  experimental finding~\cite{Bhetuwal:2020jes}, that color transparency does not occur in  $(e,e'p)$ reactions with $Q^2$, up to  14.2 GeV$^2$. The result is that   the Feynman mechanism is responsible for the proton form factor at high momentum transfer.

 \section*{Acknowledgements}
This work was supported by the U. S. Department of Energy Office of Science, Office of Nuclear Physics under Award Number DE-FG02-97ER- 41014.  I thank the organizers R. Dupre, D. Dutta, M. Sargsian and M. Strikman  of the 2021  workshop on ``The Future of Color Transparency and Hadronization Studies at Jefferson Lab and Beyond" for re-stimulating my interest in this subject. I thank S. J. Brodsky for making this paper necessary.  
%

\end{document}